\documentclass[twocolumn,pra,aps,showpacs]{revtex4}
\usepackage{epsfig}
\usepackage{hyperref}

\def\be{\begin{equation}}
\def\ee{\end{equation}}
\def\ba{\begin{eqnarray}}
\def\ea{\end{eqnarray}}

\begin{document}  
\title{Pendulum View of The Schr\"odinger Equation}
\author{Biao Wu}
\affiliation{Department of Physics, The University of Texas, Austin,
Texas 78712}
\affiliation{Condensed Matter Sciences Division, Oak Ridge National
Laboratory, Oak Ridge, Tennessee 37831}
\date{June 25, 2003}

\begin{abstract}
A transformation is found between the one dimensional Schr\"odinger equation 
and a pendulum problem. It is demonstrated how to construct exact solutions 
with the resulted pendulum equation.  The relation of this transformation
to the Zakharov-Shabat equations is pointed out. 
\end{abstract}

\maketitle
The one dimensional (1D) Schr\"odinger equation is very familiar to physicists,
most of whom have learned  quantum mechanics by first solving 
this equation for some simple potentials often used in textbooks\cite{landau}. 
This 1D equation also finds broad applications in real-world physics
with some interesting and popular potentials, such as harmonic, double-well, 
and Morse\cite{morse}. Additionally, because of its underlying mathematical structures, 
it is widely exploited in mathematical physics\cite{ge,samsonov}.

I present here a transformation that provides a new view for  the 1D Schr\"odinger 
equation. This transformation turns the 1D equation into a differential equation
describing a damped and driven pendulum, another well-known problem to physicists.
Analysis shows that, corresponding to the bound eigenstates of 
the Schr\"odinger equation, the pendulum equation has a set of special solutions, 
which are referred to as critical solutions. The number of nodes of a
bound eigenstate is related to the winding number of its corresponding pendulum motion. 
I also demonstrate how to construct exact solutions with the pendulum equation,
and reveal the relation of this transformation to 
the Zakharov-Shabat (ZS) equations\cite{zs}.

The one dimensional Schr\"odinger equation is 
\be
\label{eq:schr}
-{d^2 \over d x^2}\psi(x)+V(x)\psi(x)=\lambda^2\psi(x)\,,
\ee
where the units are chosen such that $2m=1$ and $\hbar=1$ for simplicity. Writing the 
eigen-energy $E$ as $\lambda^2$ is a choice of convenience; there is no loss of generality  
since the spectrum of a Schr\"odinger equation is bounded from below.
Introduce the transformation, 
\be
\label{eq:psi2ra}
\psi(x)=\sqrt{\rho(x)}\sin[\alpha(x)/2]\,,
\ee
where $\rho$ and $\alpha$ are real. They are different from the amplitude
and phase in the Madelung transformation where one writes $\psi=\sqrt{r}e^{i\varphi}$.
To avoid confusion, here $\rho$ is called  magnitude and $\alpha$  angle.

Straightforward calculations show that the transformation turns the Schr\"odinger 
equation into two equations,
\ba
d\alpha(x)/d x &=&
2\,\lambda-2\,A(x)\,\sin\alpha(x)\,,
\label{eq:pend}\\
\label{eq:amp}
d\rho(x)/d x&=&2\,A(x)\,\rho(x)\,\cos\alpha(x)\,.
\ea
The function $A(x)$ is related to the potential $V(x)$ as 
\be
\label{eq:pont}
V(x)=A^2(x)-dA(x)/ dx\,,
\ee 
which is called Riccati equation\cite{riccati}.

\begin{figure}[!htb]
\centerline{\includegraphics[height=4.0cm]{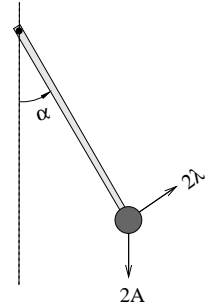}}
\caption{Schematic of a pendulum described by Eq.(\ref{eq:pend}). 
It experiences a ``gravity'' $2A$ and is driven by a constant force $2\lambda$.}
\label{fig:pendulum}
\end{figure}  
Three important and interesting observations follow  immediately. First, Eq.(\ref{eq:pend}) 
is self-contained, that is, the evolution of the angle $\alpha$ does not
depend on the magnitude $\rho$. Secondly, with the addition of $\mu d^2\alpha/dx^2$,
Eq.(\ref{eq:pend}) is in fact an equation describing a driven and damped 
pendulum (see Fig.\ref{fig:pendulum}). For this pendulum, the mass is zero ($\mu=0$), 
the driving force $\lambda$ is constant, the damping comes from a friction proportional 
to velocity, and the fictitious gravity is not constant as depicted by the force 
function $A(x)$. Thirdly, there is a symmetry in Eq.(\ref{eq:pend}): if $\alpha(x)$ is 
a solution for a given $\lambda$, then $-\alpha(x)$ is also a solution for $-\lambda$.  
As a result, I will concentrate only on the case $\lambda>0$ from now on.

For certain types of potentials (e.g., well-shaped),
the Schr\"odinger equation (\ref{eq:schr}) can have a discrete set of eigenvalues
($E_n=\lambda_n^2$, $n=0,1,2,\cdots$). The corresponding eigenstates (bound states) are 
of great importance in physics. It is interesting to know what kind of pendulum
motions these bound states have become as the result of transformation (\ref{eq:psi2ra}).
To find it out, I shall first consider one general class of finite 
force functions $A(x)$ that satisfy 
\be
\label{eq:well}
A(x)\le 1\hspace{0.6cm}{\rm and}\hspace{0.6cm}A(|x|\rightarrow \infty)=1\,.
\ee
According to Eq.(\ref{eq:pont}), these force functions lead 
to well-shaped potentials that are constant at the infinite  boundaries
and have ``dips'' in the intermediate region. 
More general cases, where the function $A$ may diverge at infinities,
will be examined later with examples.

For this class of force functions, the pendulum 
experiences a constant ``gravity'' at the boundary regimes $|x|\rightarrow \infty$, 
where the equation of motion becomes
\be
\label{eq:pendc}
d\alpha/dx=2\lambda-2\sin\alpha\,.
\ee
It has fixed point solutions when $\lambda<1$. They are
\be
f_s(\lambda)=2N\pi+\arcsin\lambda\,,~f_u(\lambda)=(2N+1)\pi-\arcsin\lambda\,,
\ee
where $N$ is an integer. All other solutions of Eq.(\ref{eq:pendc}) approach
asymptotically to $f_s$. It implies that when $\lambda<1$, the pendulum motion 
described by Eq.(\ref{eq:pend}) must start at a fixed point ($x\rightarrow -\infty$) 
and end at a fixed point ($x\rightarrow \infty$). There are four possibilities:
$f_s\rightarrow f_u$, $f_s\rightarrow f_s$, $f_u\rightarrow f_s$,
and $f_u\rightarrow f_u$.  Only the first type are the solutions 
(named as critical solutions) that correspond to the bound states of the 
Schr\"odinger equation since all the 
others have,  according to Eq.(\ref{eq:amp}), divergent $\rho$ at infinity, which 
violates the boundary conditions.

The critical pendulum solutions exist only for a discrete set of $\lambda$ that 
corresponds to the discrete eigenvalues ($E_n=\lambda_n^2$)of the Schr\"odinger 
equation (\ref{eq:schr}). 
This point can be appreciated by noting that $f_s$ is a stable fixed point and $f_u$ 
is an unstable one. Given a force function $A(x)$ and a  driving force $\lambda$, 
the pendulum motion starting at $f_s$ almost always comes back, stopping at $f_s$ since
the stable fixed point $f_s$ acts as an attractor.
In contrast, the unstable fixed point $f_u$ behaves like a repulser, 
where the pendulum would be pushed away by any small disturbance. Therefore,
only for some special values $\lambda_n$'s, the pendulum
is lucky enough to end precisely at $f_u$. In this way, the eigenvalue
problem of the 1D Schr\"odinger equation has become finding the special 
values $\lambda_n$'s in the pendulum equation (\ref{eq:pend}), where 
the critical solutions, $f_s\rightarrow f_u$, exist.

It is well known that the wave function $\psi(x)$ is oscillatory,
vanishing at positions called nodes. The transformation (\ref{eq:psi2ra}) clearly 
reflects this oscillatory nature. Furthermore, it establishes a link between the 
number of nodes of the wave function $\psi(x)$ and the number of rotations made 
in its corresponding pendulum motion: For a pendulum making $n$ 
complete rotations in a given motion, the corresponding wave function has $n$ nodes. 
 Formally, the number of rotations is called winding number, defined 
for a pendulum solution $\alpha(x)$ as
\be
W(\lambda)=\{\alpha(x\rightarrow \infty)-\alpha(x\rightarrow -\infty)\}/2\pi\,.
\ee
For a given force function $A(x)$, only winding numbers $W$ 
for solutions starting at $f_s$ are considered
for the sake of relevance and interest. As already discussed, for almost 
all values of $\lambda$, the pendulum in such motions always comes back to $f_s$ at the end. 
This means that the winding number $W(\lambda)$ is mostly an integer 
as a function of $\lambda$. Only at the special values $\lambda_n$'s,  
the winding number becomes non-integer.

\begin{figure}[!htb]
\begin{center}
\includegraphics[width=6.5cm]{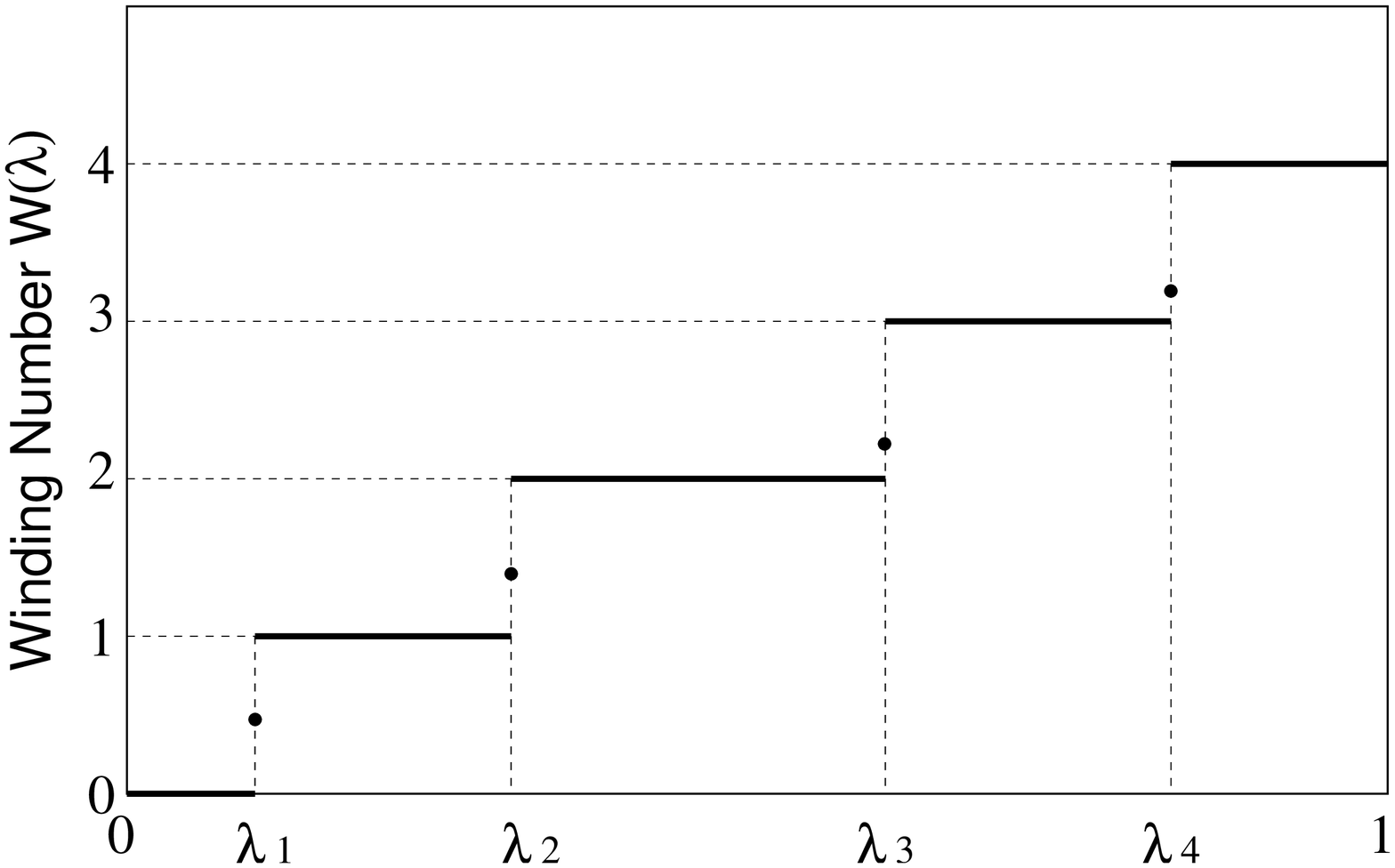}
\end{center}
\caption{Winding numbers of pendulum motions as a function
of $\lambda$. Dots denote the winding numbers at $\lambda_n$'s.}
\label{fig:step}
\end{figure}

Further analysis shows that the winding number $W(\lambda)$ should behave as depicted
in Fig.\ref{fig:step}: Starting at zero, the winding number is an 
increasing step function that jumps by one at $\lambda_n$'s.
Note that the jump at $\lambda_n$ is defined as $W(\lambda_n^{+})-W(\lambda_n^{-})$.
This conclusion is drawn on two facts. First, as shown in Appendix A,
one can prove that $W(\lambda)$ an  increasing function, that is, 
$W(\lambda^{\prime})\ge W(\lambda)$ when $\lambda^{\prime}>\lambda$ for a given $A(x)$.
Since it is integer for all values of $\lambda$ except the discrete
set, it immediately follows that it is an increasing step function. Second,
we recognize that the jump at $\lambda_n$  can only be one as dictated by
the famous oscillation theorem\cite{landau,hilbert}.
The theorem says that the wave function $\psi_n(x)$ for the $n$th bound
states has exactly $n$ nodes. It infers that the node number increases only by one
between two neighboring bound states; consequently, the winding number 
 jumps only by one at each step. 

This behavior of the winding number can be used
to estimate the number of bound states for well-shaped potentials. 
Given a  force function satisfying Eq.(\ref{eq:well}), one integrates numerically
or by other means the pendulum equation (\ref{eq:pend}) at $\lambda=1$
to find the winding number $W$, which is just the number of bound states for
the corresponding potential $V(x)$. 

\begin{figure}[!ht]
\begin{center}
\includegraphics[width=7.0cm]{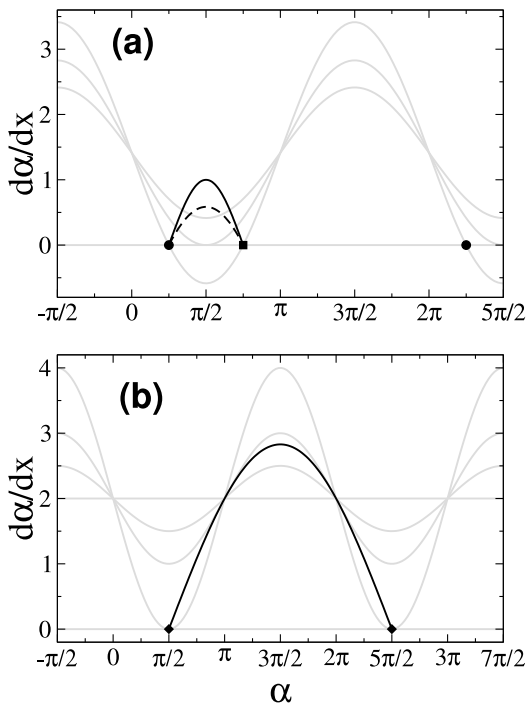}
\end{center}
\caption{Phase space trajectories of the  pendulum  as described in
 Eq.(\ref{eq:pend}) at ({\bf a}) $\lambda=1/\sqrt{2}$; ({\bf b}) $\lambda=1$.
Gray lines are for constant force functions $A$ while  dark lines
for a changing force function $A$.  Solid circles stand for the stable fixed points,
 squares for the unstable fixed points, and diamonds for the fixed points at $\lambda=1$.
}
\label{fig:phase}
\end{figure}

One general application of Eq.(\ref{eq:pend}) is to construct
exact solutions. In its phase space shown in Fig.\ref{fig:phase},
the critical solutions are represented by curves connecting
a pair of fixed points, $f_s(\lambda)$ and $f_u(\lambda)$, for a given $\lambda$.
Accordingly, the curves  are called critical curves, which are
plotted as dark lines in Fig.\ref{fig:phase}.
The following shows with examples  how to construct exact solutions from critical curves.

In one example, a critical curve is chosen for $\lambda=1/\sqrt{2}$.
Plotted as the solid dark line in Fig.\ref{fig:phase}(a), the curve
is mathematically described by $d\alpha/dx=\sin(2*\alpha-\pi/2)$. 
After integrating it for $\alpha(x)$, one can calculate out the force 
function $A$ with Eq.(\ref{eq:pend}), 
\be
A=\frac{\sqrt{2}\cosh(2x)-1}{2\cosh(x)\sqrt{\cosh(2x)}}\,.
\ee
According to Eq.(\ref{eq:pont}),  the related potential is
\be
V=\frac{\sinh(x)-(\sqrt{2}+1)\sinh(3x)}{2\sqrt{2}\cosh^2(x)\cosh^{3/2}(2x)}+
\frac{[1-\sqrt{2}\cosh(2x)]^2}{4\cosh^2(x)\cosh(2x)}\,,
\ee
which has an eigenvalue at $E_0=\lambda_0^2=1/2$. By calculating
the winding number at $\lambda=1$, one knows that this is
the only discrete eigenvalue. The corresponding eigenfunction
can be found with  Eqs.(\ref{eq:pend},\ref{eq:amp}); it is
\be
\psi_0=\sqrt{\frac{[2\cosh(2x)]^{1/2}}{(2\cosh x)^{1+\sqrt{2}}}}
\sin\Big(\frac{\pi+\arctan[\sinh(2x)]}{4}\Big)\,.
\ee

In another example, the critical curve is 
given by $d\alpha/dx=-2\bar{\lambda}+2\sin\alpha$ 
(the dark dashed line in Fig.\ref{fig:phase}(a)), which connects
a pair of fixed points, $f_s(\bar{\lambda})$ and $f_u(\bar{\lambda})$ for any
$\bar{\lambda}<1$.
The same steps as in the first example lead to the potential,
\be
\label{eq:pot1}
V(x)=1-2(1-\bar{\lambda}^2)/\cosh^2(\sqrt{1-\bar{\lambda}^2}x)\,.
\ee
This potential has only one discrete eigenvalue at $\bar{\lambda}^2$; it
is a special case of a well-known potential that has exact solutions\cite{landau}.
Other critical curves, especially, the ones allowing quadrature integration of
$d\alpha/dx=f(\alpha)$, can
be chosen and used to construct potentials for a predetermined eigenvalue.

Similarly, one can also construct a potential that 
has a desired number of discrete eigenvalues. The basic strategy is to use 
the curves that connect a pair of fixed points for $\lambda=1$, 
which are at $f_1=2N_1\pi+\pi/2$ and $f_2=2N_2\pi+\pi/2$.
The potential so constructed will have exactly $N_2-N_1$ bound states.
The example is $d\alpha/dx=2\sqrt{2}\sin(x/2-\pi/4)$ 
(the dark curve in Fig.\ref{fig:phase}(b)), which connects
$f_1=\pi/2$ and $f_2=5\pi/2$. 
With the similar steps as in the above two examples, one finds
\be
\label{eq:pot2}
V(x)=\frac{\cosh(2\sqrt{2}x)-4\sinh(\sqrt{2}x)+1}{\cosh(2\sqrt{2}x)+
4\sqrt{2}\cosh(\sqrt{2}x)+5}\,.
\ee
One can verify numerically that this potential has, indeed, only 
one eigenvalue,  approximately at $\lambda^2= 0.5$.

So far, only the cases satisfying Eq.(\ref{eq:well}) are considered. However,
many concepts, such as critical solutions and winding numbers, are general and
applicable to other cases, where $A$ may be divergent at infinities.
I demonstrate it with an example, $A(x)=x$, which corresponds to a harmonic potential 
$V(x)=x^2-1$. For this case, the pendulum equation becomes 
\be
\label{eq:pendh}
d\alpha/dx=2\lambda-2x\sin\alpha\,.
\ee
Consider solutions starting at $\alpha=\pi$ as one can check that only
such solutions have chance to satisfy boundary conditions for bound states. 
Through numerical calculations, 
one finds that for almost any value of $\lambda$, 
these solutions end asymptotically at  $\alpha=2(n+1)\pi$ ($n=0,1,2,\cdots$),
and the winding number jumps by one approximately at $\lambda_n=\sqrt{2n}$ (note
that $E_n=2n$ is an eigenvalue of the harmonic potential $V(x)=x^2-1$). 
Numerical calculations show, at these discrete values $\lambda_n$s, the pendulum stays close
at $\alpha=(2n+1)\pi$ for some time then strays away, apparently
due to numerical errors. These special pendulum solutions at $\lambda_n$ are
exactly the critical solutions mentioned before; they correspond to the bound states
of the Schr\"odinger equation. 
Detailed analysis of the asymptotic behavior of Eq.(\ref{eq:pendh})
indicates that $\alpha=(2n+1)\pi$ acts like a repulser while $\alpha=2(n+1)\pi$ like
an attractor even though they are not fixed points in general. 
Finally, look at one special case,
$\lambda_0=0$, where the critical solution
is $\alpha(x)=\pi$, accompanied by 
$
\rho=e^{-x^2}\,.
$
Combined according to Eq.(\ref{eq:psi2ra}), they produce exactly the ground state 
in the harmonic potential $V(x)=x^2-1$.

There is another similar transformation that yields the same equations as 
Eqs.(\ref{eq:pend},\ref{eq:amp}).  It is 
\be
\label{eq:psi2ra2}
\psi(x)=\sqrt{\rho(x)}\cos[\alpha(x)/2]\,.
\ee
However, the potential and the force function are related differently
\be
\widetilde{V}(x)=A(x)^2+dA(x)/dx\,.
\ee
For a given $A(x)$, if the pendulum equation Eq.(\ref{eq:pend}) has critical
solutions at  $\lambda_n$'s, then both potentials $V(x)$ and $\widetilde{V}(x)$ have 
the same discrete eigenvalues, $\lambda_n^2$'s. In short, 
the two potentials $V(x)$ and $\widetilde{V}(x)$ have the same eigenvalues.
Note that this conclusion is true only when
Eq. (\ref{eq:pend}) has critical solutions for a force function $A(x)$.
For example, it is not true when $A(x)$ is a step function, for which
Eq.(\ref{eq:pend}) has no critical solutions.

Finally, I point out that the transformations  
Eqs.(\ref{eq:psi2ra},\ref{eq:psi2ra2}) are related to the ZS equations\cite{zs}. 
In Ref.\cite{inter}, a transformation is found reducing the ZS equations to a
Schr\"odinger equation; in Ref.\cite{dsoliton}, another transformation is
discovered turning the ZS equations into equations similar to 
Eqs.(\ref{eq:pend},\ref{eq:amp}).
Their combination produces these two transformations  
Eqs.(\ref{eq:psi2ra},\ref{eq:psi2ra2}), whose
generalized versions are given in Appendix B.

The author thanks Junren Shi for helpful discussion.
This work was supported by the NSF and the LDRD of ORNL, managed by UT-Battelle
for the USDOE (DE-AC05-00OR22725). 
\appendix
\section{Winding number $W(\lambda)$}
\noindent{\it For a given force function $A(x)$, $W(\lambda^{\prime})\ge W(\lambda)$
if $\lambda^{\prime}>\lambda$}.\\

\noindent{\bf Proof:}
Assume the opposite, $W(\lambda^{\prime})< W(\lambda)$.
Since the difference is at least one, the pendulum must make more
rotations at $\lambda$ than at $\lambda^{\prime}$. Note also
that at $x\rightarrow -\infty$ one has 
$\alpha(x,\lambda^{\prime})>\alpha(x,\lambda)$. These two
facts infer that there must exist a point $\tilde{x}$, where the two pendulum
motion functions, $\alpha(x,\lambda^{\prime})$ and $\alpha(x,\lambda)$ 
cross each other for the first time. At this crossing position, $x=\tilde{x}$,
one should have $\frac{d\alpha}{dx}(\tilde{x},\lambda^{\prime})
<\frac{d\alpha}{dx}(\tilde{x},\lambda)$. However, this contradicts the fact,
\be
\frac{d\alpha}{dx}(\tilde{x},\lambda^{\prime})
-\frac{d\alpha}{dx}(\tilde{x},\lambda)=2(\lambda^{\prime}-\lambda)>0\,.
\ee
So, the assumption is not true. We have $W(\lambda^{\prime})\ge W(\lambda)$.

\section{Transformations of Zakharov-Shabat equations}
The Zakharov-Shabat (ZS) equations are 
\ba
\displaystyle id U_1(x)/d x +  
\phi(x)U_2(x) &=& \lambda U_1(x)\,,\\
\displaystyle id U_2(x)/d x -  
\phi^*(x)U_1(x) &=& -\lambda U_2(x)\,,
\ea
where the complex function $\phi(x)$ acts as a potential. 
We choose to write 
$ 
\phi(x)=A(x)e^{iS(x)}
$, 
where $A(x)>0$ and $S(x)$ are real. 

\noindent{\it Schr\"odinger equation.--} Introduce a transformation
\be
\label{eq:zs2schr}
\psi_{\pm}(x;\lambda)={U_1(x)e^{-iS(x)/2}
\pm i\,U_2(x)e^{iS(x)/2}\over \sqrt{2\lambda+S^\prime(x)}}\,,
\ee
where $S^\prime=dS/dx$. 
By straightforward computation, one can verify that $\psi_{\pm}$ satisfy the 
Schr\"odinger equation
\be
\label{eq:schr1}
-d^2 \psi_\pm/d x^2+V_{\pm}(x)\psi_\pm=
(\lambda+S^{\prime}/2)^2\psi_\pm\,,
\ee
where
\be
V_{\pm}(x)=(A^2\pm A^\prime)\pm{AS^{\prime\prime}
\over 2\lambda+S^{\prime}}-{2S^{(3)}(\lambda+S^\prime)-
3(S^{\prime\prime})^2\over 4(2\lambda+S^{\prime})^2}\,.
\ee
In the special case $S(x)={\rm constant}$, the above equation becomes
\be
\label{eq:schr2}
-d^2 \psi_\pm/ d x^2+V_{\pm}(x)\psi_\pm=
\lambda^2\psi_\pm\,,
\ee
with much simpler potentials  
\be
V_{\pm}(x)=A^2\pm A^\prime\,.
\ee 
These two potentials $V_\pm$ are $\widetilde{V}$ and $V$ used in the main text.\\

\noindent{\it Pendulum equation.--} For the discrete eigenvalues, 
the ZS equations have $|U_1|=|U_2|=\sqrt{\rho}$. This allows one to introduce another 
transformation with
\be
\label{eq:zs2pend}
U_1=\sqrt{\rho}\,e^{i\,\theta_1}\,,~~~~~U_2=\sqrt{\rho}\,e^{i\,\theta_2}\,.
\ee
Substituting it into the ZS equations, one obtains
\ba
d\rho/dx&=&2\,A\,\rho\sin(\theta-S)\,,\\
d\theta/dx&=&-2\lambda+2\,A\,\cos(\theta-S)\,,
\ea
where $\theta=\theta_1-\theta_2$.  The above two equations become 
Eqs.(\ref{eq:pend},\ref{eq:amp}) when $S(x)=0$ with $\alpha=\pi/2-\theta$.\\

The combination of Eq.(\ref{eq:zs2schr}) and Eq.(\ref{eq:zs2pend})
results in the transformations Eqs.(\ref{eq:psi2ra},\ref{eq:psi2ra2}).



\begin{thebibliography}{99}
\bibitem{landau}L.D. Landau and E.M. Lifshitz, {\it Quantum mechanics} 
(Pergamon Press, New York, 1977).

\bibitem{morse}P.M. Morse, Phys. Rev. {\bf 34}, 57 (1929).

\bibitem{ge}Mo-Lin Ge {\it et al.}, 
Phys. Rev. A {\bf 62}, 052110 (2000).

\bibitem{samsonov}B.F. Samsonov, Phys. Lett. A{\bf 263}, 274 (1999),
B. Bagchi and A. Ganguly, math-ph/0302040.

\bibitem{zs} V.E. Zakharov and A.B. Shabat, Zh. Eksp. Teor. Fiz. {\bf 64},
1627 (1973)[Sov. Phys. JETP {\bf 37}, 823 (1974)].


\bibitem{riccati}Einar Hille, {\it Lectures on Ordinary
differential Equations} (Addison-Wesley, Reading, 1969).

\bibitem{hilbert}R. Courant and D. Hilbert,
\textit{Methods of mathematical physics} (Interscience, New York, 1953).

\bibitem{inter}W.M. Liu, B. Wu, and Q. Niu, Phys. Rev. Lett. {\bf 84},
2294 (2000).

\bibitem{dsoliton}B. Wu, J. Liu, and Q. Niu, 
Phys. Rev. Lett. {\bf 88}, 034101 (2002).

\end{thebibliography}
\end{document}